\documentstyle[12pt,amsmath,amssymb]{article}
\def\beq{\begin{equation}}
\def\eeq{\end{equation}}

\def\al{\alpha}

\def\ga{\gamma}

\def\de{\delta}
\def\De{\Delta}
\def\si{\sigma}

\def\La{\Lambda}

\def\sq{\sqrt}

\def\l{\left (}
\def\r{\right )}
\def\fr{\frac}
\def\la{\label}
\def\hs{\hspace}

\def\ov{\overline}

\begin{document}
\begin{titlepage}

\begin{center}   
{\Large\bf 
Varying Couplings From Orbifold GUTs}  
\end{center} 
\vspace{0.5cm} 
\begin{center} 
{\large Filipe \hs{-0.07cm}Paccetti \hs{-0.07cm}Correia$^{a}$\footnote{E-mail address: 
F.Paccetti@ThPhys.Uni-Heidelberg.DE} , 
{}Michael \hs{-0.07cm}G. \hs{-0.07cm}Schmidt$^{a}$\footnote{E-mail address: 
M.G.Schmidt@ThPhys.Uni-Heidelberg.DE} , 
{}Zurab \hs{-0.07cm}Tavartkiladze$^{a, b}$\footnote{E-mail address: 
Z.Tavartkiladze@ThPhys.Uni-Heidelberg.DE} 
} 
\vspace{0.5cm}

$^a${\em Institut f\"ur Theoretische Physik, Universit\"at Heidelberg,
Philosophenweg 16,\\
 D-69120 Heidelberg, Germany \\

$^b$ Institute of Physics, 
Georgian Academy of Sciences, Tbilisi 380077, Georgia\\

} 

\end{center}

\vspace{1.0cm}

\begin{abstract}

We discuss the variation of gauge couplings in time in the framework of orbifold
constructions, due to a change of the extra compact
dimension's size. Models with gauge coupling unification allow to
estimate the variation of the strong coupling constant $\al_3$ and to relate it to a variation of $\alpha_{em}$.  The extra-dimensional
construction turns out to be crucial for the model to be compatible with
data. Within the presented 5D scenarios,  the tower of KK states significantly affects
gauge coupling running, leads to low scale unification, and provides
a suppression of the $\al_3(M_Z)$ variation. 

\end{abstract}

\end{titlepage}


\section{Introduction}

The recent measurements \cite{webb} of quasar absorption lines hint to a
variation of the fine structure constant (at redshift range $0.2<z<3.7$) with
\beq
\fr{\de \al }{\al }=\fr{\al_0-\al }{\al}=(-0.57\pm 0.10)\cdot 10^{-5}~,
\la{vardata}
\eeq
where $\al_0$ is the value of the fine structure constant at late
cosmological times ($z=3.5$), while $\al $ corresponds to the value at
present time.
These data renewed an interest in the old issue of the possible variation of the nature 'constants' \cite{dirac}, \cite{beker}, 
triggered the construction of models 
\cite{dvali}-\cite{dine}, which suggest theoretical frameworks
for this
phenomenon and have reactivated discussions \cite{disc}. 
The various implications of such  models are
interesting and deserve detailed studies.

The issue of the variation of the fine structure constant got a new insight
within
GUTs \cite{old and new}-\cite{dine}. Since above the unification
scale the gauge sector is
described by one unified gauge coupling, the variation of $\al $ 
also directly determines the time dependence of the QCD coupling $\al_3$. This 
would affect standard big bang nucleosynthesis and high-redshift
quasar absorption lines much more than only a
variation of the fine structure constant. 
In the realistic
versions of GUT the $\al_3$ variation contradicts \cite{dine} the available data and only
some specific extension may save the situation.  This
issue can be considered as a new selection rule in building 
realistic GUTs, even without having a dynamical mechanism for the variation of
couplings.
Of course, it is also desirable
to have a theoretical understanding of this
variation. Within superstring and/or
higher dimensional constructions the four dimensional couplings often
depend on VEVs of some scalar fields coming from
higher dimensions.
It is then natural to have time dependent couplings, because on the 4D level
they are related to some dynamical fields \cite{dvali}.

In this paper we consider 5D SUSY scenarios compactified on an orbifold.
Orbifold constructions have recently gained much attention as they
present a 
framework where typical problems of 4D GUTs, such as the 
Doublet-Triplet  splitting problem and
strong baryon number violation, are easily avoided
\cite{kawa}-\cite{us}. The GUT symmetry breaking also can be realized in a
rather economical way.
Interestingly enough, in this setting the source of coupling
variation can be the change of the size $R$ of the compact fifth 
dimension. This
framework in addition to a variation of $\al $ can also give a variation
of
$\al_3$ (and of other couplings which are expressed by 5D parameters 
and $R$) without gauge coupling unification.
Having a 
unification condition at a certain scale (interpreted as the GUT scale
$M_G$), 
allows one to
relate $\fr{\de \al_3}{\al_3}$ to $\fr{\de \al }{\al }$ at low
energies. The value of $\fr{\de \al_3(M_Z)}{\al_3(M_Z)}$ depends on the
intermediate thresholds contributing between the scales $M_Z$ and $M_G$. As we
will see, with a compactification scale $\mu_0=\fr{1}{R}\simeq M_G$, the
picture does not differ from the one of usual 4D GUTs, having similar
implications. However, with $\mu_0\ll M_G$, the KK
states will participate in renormalization. The latter turn out 
to play an essential role
for 
$\fr{\de \al_3(M_Z)}{\al_3(M_Z)}$ suppression compared to 
4D GUT scenarios (a different possibility for the suppression of
$\fr{\de \al_3}{\al_3}$ was discussed in the $1^{\rm st}$ ref. of
\cite{more}). 
The suppression becomes stronger when many KK
states appear
below $M_G$, making the scenario compatible with observations without
affecting significantly the predictions of nucleosynthesis.
Let us note that a
large number of KK states, 
as was pointed out in \cite{dines}, can be crucial also 
in lowering the unification scale down to a few TeV. 
This was confirmed 
with explicit models in \cite{kakushadze}, \cite{lowscale}, \cite{us}.
A low unification scale makes a model testable in the next generation
of high energy experiments \cite{ant}.



\section{Some Bounds from Experimental Data}

Several authors have studied the possibility that the variation in the 
fine structure constant $\alpha$ is the consequence of the 
(more fundamental) 
variation of the unified coupling $\alpha_G$ either at a fixed 
GUT scale $M_G$ \cite{old and new} or at a varying one 
\cite{calmet,dine}. In view 
of this, it is clear that the variation in $\alpha$ should be accompanied
by variations in the other gauge couplings, but possibly also 
by variations in Yukawa couplings and mass scales such as the 
SUSY and electroweak scales\footnote{The variation of a mass scale 
$m$ is naturally to be understood as measured in 
units of some non-varying mass scale $M$, 
i.e. $\delta m=\delta s \cdot M$ where $s=m/M$. In the next section we will choose $M=M_G$.}. 
As we will see, it turns out that the knowledge of 
the mechanism which is behind the variation of the fundamental
 constants is at least as important as the knowledge of the 
model of unification.

The variations of various observables are constrained by observational
data and if we trust these data one can check various scenarios. 
It is beyond the scope of this paper to consider precise mechanisms of (N=1) 
SUSY breaking, EW symmetry breaking nor do we precise the way $M_{Pl}$ varies. 
We will therefore be forced to introduce several dimensionless quantities 
($\rho,~\rho_a,~\bar{\rho}_c,\dots$) to parameterize our ignorance. 
In fact, in the kind of models we consider here, these make only sub-leading (but non-negligible) contributions to the variations of the observables we are interested in.

A variation in the strong coupling
implies a variation in the QCD scale $\Lambda_c$. 
The latter can be roughly expressed through $M_Z$ as 
\beq
\La_c=M_Z{\rm exp}\l -\fr{2\pi}{9\al_3(M_Z)}\r~,
\la{QCDslace}
\eeq 
where thresholds are neglected.
Therefore,
\begin{equation}
\frac{\delta\Lambda_c}{\Lambda_c}
=\l \fr{2\pi}{9}\al_3^{-1}+\ov{\rho }_c\r \fr{\de \al_3}{\al_3}~,
\la{varQCD}
\end{equation}
where $\ov{\rho }_c$ includes possible threshold contributions in the range
$M_Z-\La_c$ and is expected to have a sub-leading effect.
The resulting variation in the masses of the hadrons is constrained by 
some cosmological observables such as the abundance of elements
\cite{berg} 
and the high redshift quasar
absorption lines \cite{lines1, lines2}.

The ${}^4{\rm He}$ abundance $Y_4$ is expressed as
$Y_4=\fr{2n_n/n_p}{1+n_n/n_p}$, where 
$n_n/n_p\simeq 0.8\cdot e^{-Q/T_D}$($\simeq 1/7$) 
is the ratio of neutron to proton density
at the time of nucleosynthesis, with a decoupling temperature
$T_D\simeq 0.8$~MeV and mass difference $Q=m_n-m_p=B+C\al \La_c\simeq
1.29$~MeV ($B=2.05$~MeV, $C\sim -1$).
$T_D$ is determined by the expansion rate, $T_D\sim (M_{Pl}G_F^2)^{-1/3}$. Having
a variation of couplings/scales, $M_{Pl}$ and $G_F$ can also vary 
(as we already pointed out, one mass scale, not necessarily $M_{Pl}$, 
can be fixed as a reference scale since only the 
variation of dimensionless quantities has a physical meaning). 
Without detailed knowledge of the variation law 
one can parameterize our ignorance with
$\fr{\de (M_{Pl}G_F^2)}{M_{Pl}G_F^2}\equiv\rho \fr{\de \al }{\al
}$\footnote{As it
turns out, for the considered scenarios $\rho $ will not
play a significant role. This allows one to carry out the analysis without knowledge of 
$\rho $.} and therefore 
$\fr{\de T_D}{T_D}=-\rho\fr{\de \al }{3\al }$. Taking into account all this,
one obtains
$$
\fr{\de Y_4}{Y_4}=\fr{B-Q}{T_D}\fr{1}{1+n_n/n_p}\l 
\fr{\de \al }{\al }+\fr{\de \La_c}{\La_c}+\fr{Q}{B-Q}\fr{\de T_D}{T_D}\r
$$
\beq
{}\simeq 0.8 (1+{\cal R}-0.6 \rho )\fr{\de \al }{\al }~,
\la{4Heabund}
\eeq
where we defined
\beq
\fr{\de \La_c}{\La_c}={\cal R}\fr{\de \al }{\al }~.
\la{defs}
\eeq
The value of $\fr{\de Y_4}{Y_4} $, at times corresponding to 
$z\sim 10^{10}$, is constrained \cite{berg}

\beq
\fr{\de Y_4}{Y_4} = (-5.6 \pm 7.2) \cdot 10^{-3}~.
\la{Y4lim}
\eeq
If a time independent $\fr{\dot{Y}_4}{Y_4}$ is assumed, then using in
(\ref{4Heabund}) the value (\ref{vardata}) (which corresponds to
$z=0.5-3.5$) we can verify that (\ref{Y4lim}) is indeed satisfied (unless 
$|{\cal R}-0.6\rho| \stackrel{>}{_\sim }10^3$, which is never realized as
we will see below). The situation can be drastically changed for a time
dependent $\fr{\dot{Y}_4}{Y_4}$, since integration in a range of large
redshifts can cause a strong enhancement in $\fr{\de Y_4}{Y_4}$. However,
since the present data doesn't have any information about the law by which 
$\fr{\dot{Y}_4}{Y_4}$ may vary, we are unable to
judge this situation.


Other observables that one can use to constrain any model are 
obtained from high precision observations  of 
quasar absorption lines \cite{lines1}:
\begin{equation}
X\equiv \alpha^2 g_p \frac{m_e}{m_p},\quad \textup{with } 
\frac{\delta X}{X}=(0.7\pm 1.1)\times 10^{-5}\quad (z=1.8),
\la{Xlim}
\end{equation}
and measuring the wavelengths of molecular hydrogen transitions in the
early universe \cite{lines2}
\begin{equation}
Y\equiv \frac{m_p}{m_e},\quad \textup{with }\frac{\delta Y}{Y}=
(3.0 \pm 2.4)\times 10^{-5}\quad (z=3).
\la{Ylim}
\end{equation}
Using
\begin{equation}
           \frac{\delta X}{X}=2\frac{\delta \alpha}{\alpha} -\frac{\delta Y}{Y},
\la{conflim}
\end{equation}
(where we neglect the variation of $g_p$\footnote{See however
discussions in \cite{shuryak}.}) 
one sees that at $1\sigma$ level there is an inconsistency between 
the data (\ref{Xlim}), (\ref{Ylim}), and the measured variation
(\ref{vardata}) of
the fine structure constant. However, taking data at
$2\sigma$ level (in the spirit of ref.\cite{dine}), 
from (\ref{conflim}) it follows that the measurements are
consistent with each other. Because of this, 
to derive further bounds on ${\cal R}$, the corresponding data at $2\si $
level is used. Taking into account $m_p\sim \La_c$ and eq.(\ref{defs}),
we have $\fr{\de X}{X}=(2-{\cal R}+\rho_e)\fr{\de \al }{\al }$ and 
$\fr{\de Y}{Y}=({\cal R}-\rho_e)\fr{\de \al }{\al }$, where $\rho_e$ 
is introduced to parameterize the variation of $m_e$, $\frac{\de m_e}{m_e}=\rho_e \fr{\de \al}{\al}$.
Using
eqs.(\ref{vardata}), (\ref{Xlim}) and (\ref{Ylim}) - all at $2\si $ level - we
finally obtain \footnote{This bound is obtained from the unweighted 
data (eq.\eqref{Ylim}) of the second reference in \cite{lines2}. 
With the weighted data we would need to go to 3$\sigma$ to satisfy \eqref{conflim}. In that case the range for $({\cal R}-\rho_e)$ would be shifted.} 
\begin{equation}
-2.1 + \rho_e \stackrel{<}{_\sim} {\mathcal R} \stackrel{<}{_\sim} 4.9 + \rho_e~.
\la{constraint}
\end{equation}
To check whether this limit is satisfied for a given model or not, we 
need to know ${\cal R}$ (and $\rho_e$).

\section{The Framework}

Consider 5D SUSY models compactified on  
a circle of radius $R$ which is allowed to vary. This is the source of time
dependent couplings. We will discuss scenarios with gauge coupling
unification at $M_G$, not much below the 5D fundamental mass
scale $M_{*}$.
It is natural to assume that the dimensionfull 5D gauge coupling 
$\alpha_5^{-1}$ is close to the fundamental scale.
With $R \gg M_{*}^{-1}$ (which is indeed satisfied in the models considered
below) the 4D unified 
coupling $\al_G$ at the $M_G$ scale (and its vicinity) has the following
$R$-dependence:
\begin{equation}
\alpha_G^{-1}(M_G)\simeq 2\pi R \alpha_5^{-1}.
\la{rel45}
\end{equation}   
Apart from the change 
in $\alpha_G^{-1}$ we have a variation in the masses of the Kaluza-Klein 
states which, as we will see below, can contribute significantly (at least
for $RM_G\gg 1$) 
to the variation of gauge couplings at low energies. To estimate this
effect we should renormalize couplings from high energies down to low
scales.

The gauge coupling evolution from a high scale $\mu $ down to the 
weak scale in
one loop approximation reads:
\begin{equation}\label{eq:1}
\al_a^{-1}(M_Z)=\al_a^{-1}(\mu ) +
\frac{b_a}{2\pi}\ln{\frac{\mu }{M_Z}}+P_a+\Delta_a,
\end{equation}
with $P_a$ the contribution of the KK states (if $\mu>\mu_0=1/R$) and
$\Delta_a$ the 
contribution of other possible (logarithmic) thresholds. 
Having the unification condition $\al_a(M_G)=\al_G(M_G)$ at $M_G$, from 
the three 
equations (\ref{eq:1}) we get
\begin{equation}
\alpha_3^{-1}=\frac{12}{7}\alpha_2^{-1}-\frac{5}{7}\alpha_1^{-1}+(P_3-\frac{12}{7}P_2+\frac{5}{7}P_1)+(\Delta_3-\frac{12}{7}\Delta_2+\frac{5}{7}\Delta_1),
\la{al3}
\end{equation}
\begin{equation}
\ln\frac{M_G}{M_Z}=\frac{5\pi}{14}(\alpha_1^{-1}-\alpha_2^{-1})-\frac{5\pi}{14}\left(P_1-P_2\right)-\frac{5\pi}{14}\left(\Delta_1-\Delta_2\right),
\la{MG}
\end{equation}
\begin{equation}
\alpha_G^{-1}=\alpha_2^{-1}-\frac{1}{2\pi}\ln\frac{M_G}{M_Z}-P_2-\Delta_2,
\la{alG}
\end{equation}
where we have taken into account that 
$(b_1,~b_2,~b_3)=(\frac{33}{5},~1,~-3)$. Expressions
(\ref{al3})-(\ref{alG}) will be useful for estimating the unification picture
within specific models.

In the following,  we will consider two types of 
compactification, which give different mass spectra for KK states. With
compactification on an $S^1/Z_2$ orbifold, all states have definite $Z_2$
parity ${\cal P}=\pm 1$ under $y\to -y$ (fifth space-like dimension) and
the masses of KK states are $\fr{n}{R}$, where $n$ denotes the quantum number
in the KK mode expansion. In this case
\begin{equation}
P_a = \fr{\hat{b}_a}{2\pi}\sum_{n=1}^{N_0}\ln{\fr{M_G R}{n}},
\end{equation}
where $N_0$ is the truncation number of the KK 
tower\footnote{See ref.\cite{us} for more details.} ($N_0=[\fr{M_G}{\mu_0}]$),
and $\hat{b}_a$ is a model-dependent group theoretical 
factor.

With compactification on an $S^1/Z_2\times Z_2'$ orbifold
the bulk states have parities $(\pm 1, \pm 1)$ and $(\pm 1, \mp 1)$
with masses $\fr{2n+2}{R}$ and $\fr{2n+1}{R}$, respectively, for
the corresponding KK states. In this case we have
\begin{equation}
P_a = \fr{\ga_a}{2\pi}\sum_{n=0}^{N}\ln{\fr{M_G R}{2n+2}}+
\fr{\de_a}{2\pi}\sum_{n=0}^{N'}\ln{\fr{M_G R}{2n+1}},
\end{equation}
($N=[\fr{M_G}{2\mu_0}-1]$, $N'=[\fr{M_G}{2\mu_0}-\fr{1}{2}]$), where
$\gamma_a$ and $\delta_a$ are group-theoretical factors corresponding to
states with parities $(\pm 1, \pm 1)$ and $(\pm 1, \mp 1)$ respectively.

Now we investigate the effects of a variation $\delta R$ of $R$, the 
size of the 5th dimension. Since only the variation of the 
ratio of mass scales has physical meaning, for convenience we will 
treat $M_G$ as a fixed reference scale. 
Then other scales will be time dependent.
If the number of 
KK states is few orders of magnitude larger than the number of the other 
thresholds, then the latter will have sub-leading effect. This statement
gets more support if all matter and MSSM higgs doublets live on a brane
and have $R$ independent couplings. We discuss this at the end of
sect. 4.1.
Taking all this into account, from eqs.\eqref{rel45} and \eqref{eq:1} we
find
\begin{equation}
\al_a^{-1}\fr{\de \al_a}{\al_a}=-A_a\frac{\delta R}{R},
\la{vareq}
\end{equation}
where $\al_a$ are the couplings at $M_Z$ and
\beq
A_a=A_a^0+\rho_a~,
\la{totA}
\eeq
\begin{equation}\label{eq:A}
A_a^0\equiv \al_G^{-1}+\fr{1}{2\pi }\left\{\begin{array}{ll}
\hat{b}_aN_0 &{\rm for}~ S^1/Z_2 ~{\rm orbifold},\\
\ga_a(N+1)+\de_a(N'+1) &{\rm for}~
S^1/Z_2\times Z_2' ~{\rm orbifold}, \end{array}\right. ~
\end{equation}
We have assumed that $\al_5$ does not vary in time, i.e. the time
dependence in $\al_G$ is caused only by the variation of $R$.
In (\ref{totA}) $\rho_a$ denote the effects of the variation of the thresholds $\De_a$ in the
$M_G-M_Z$ range.
These include the effects of the variations in SUSY and EW scales. 
They depend on the origin of SUSY and EW symmetry breaking and the
mechanism of generation of fermion masses. 
Without specifying these, it is still possible to
demonstrate the effects caused by the KK states.

In addition to the three equations (\ref{vareq}) we have the relation
$\alpha^{-1}(M_Z)=(5/3)\alpha_1^{-1}(M_Z)+\alpha_2^{-1}(M_Z)$. This allows
us to eliminate $\fr{\de R}{R}$ and to express $\fr{\de \al_3}{\al_3}$ through
$\fr{\de \al}{\al}$:
\beq
\fr{\de \al_3}{\al_3}=\fr{\al_3}{\al }\fr{3A_3}{5A_1+3A_2}
\fr{\de \al}{\al }~.
\la{varal3}
\eeq
On the r.h.s. of this equation $\al_3$ is taken at $M_Z$, while $\al $
at the electron mass. The reason for the latter is that $\fr{\de \al}{\al}$ is
measured at low scales and the combination $\al^{-1}\fr{\de \al }{\al }$
is scale invariant in the leading order approximation. 
Using (\ref{varQCD}) and (\ref{totA})-(\ref{varal3}) we can express
${\cal R}$ as
\beq
{\cal R}\approx \fr{2\pi }{9\al }\fr{3A_3^0}{5A_1^0+3A_2^0}
\l 1+\fr{9\al_3}{2\pi }\ov{\rho }_c+\fr{\rho_3}{A_3^0}-
\fr{5\rho_1+3\rho_2}{5A_1^0+3A_2^0} \r ~,
\la{aprR}
\eeq
where we have assumed that $\ov{\rho}_c$, $\rho_a$ have sub-leading
effects.
{}For a given model one can estimate unification by 
using eqs.(\ref{eq:1}), calculate all $A_a$ factors, and then 
determine ${\cal R}$ from (\ref{aprR}). 
In the case that all KK states lie above
$M_G$, they are irrelevant for the running and we have
$A_a^0=\al_G^{-1}\simeq 24$. Therefore,
\beq
{\cal R}_{\rm 4D~GUT}\approx 36\l 
1+0.17~\ov{\rho }_c+0.04(\rho_3-\fr{5}{8}\rho_1-\fr{3}{8}\rho_2)\r~,
\la{R4dGUT}
\eeq  
and we
recover the result for 4D GUTs \cite{old and new}-\cite{dine} (there
was not presented expression including 
$\ov{\rho}_c$, $\rho_a$ in these papers).
The value of ${\cal R}$ in (\ref{R4dGUT}) should be compared with the
upper bound of (\ref{constraint}). In order to satisfy this bound, either
$\ov{\rho }_c$ must equal to $-5.39$ with $1\%$ accuracy or
$\rho_3-\fr{5}{8}\rho_1-\fr{3}{8}\rho_2=-22.9$ with the same accuracy (these
are nothing but fine tunings with $1\%$ accuracy). It is hard to
imagine having such cancellations. Because of this, the 4D GUTs turn out
to be inconsistent \cite{dine} with data.

To compare the magnitude of the variation of the strong coupling in KK models with the one in the 4D GUTs, we introduce the quantity 
\begin{equation}\label{eq:kappa}
\kappa \equiv \frac{8A_3^0}{5A_1^0+3A_2^0}~.
\end{equation}
which measures the suppression of $\fr{\de \al_3}{\al_3}$ (see
(\ref{varal3})) compared to the one in 4D GUTs.
It is clear that in the absence of KK states we obtain 
$\kappa\sim 1$. Once more we emphasize that this is usually the case for
realistic 
4D GUTs \cite{dine}. Such a $\kappa$ is unsatisfactory as it is 
too large when compared with the constraint which (taking into
account (\ref{aprR})) follows from 
eq.(\ref{constraint})
\begin{equation}
-0.06\stackrel{<}{_\sim } \kappa 
\l 1+0.17~\ov{\rho }_c+\fr{\rho_3}{A_3^0}-
\fr{5\rho_1+3\rho_2}{5A_1^0+3A_2^0} \r
\stackrel{<}{_\sim } 0.15~.
\la{kapconst}
\end{equation}

Inspection of eqs.(\ref{eq:A}) and (\ref{eq:kappa}) shows that the 
presence of a large number of KK modes may significantly change this 
situation, hopefully by suppressing the variation of the strong coupling. 
As we will show now, this is indeed the case for low scale unification 
models. The point is that for large $N_0$, in order to maintain the successful
prediction for $\al_3(M_Z)$, we need 
\begin{equation}
\hat{b}_3=\frac{12}{7}\hat{b}_2-\frac{5}{7}\hat{b}_1.
\end{equation}
(For the $S^1/Z_2\times Z_2'$ orbifold we must replace 
$\hat{b}_a\to \gamma_a+\delta_a$ and $N_0\to N+N'+2$). 
We then have 
$\alpha_3^{-1}\approx (12/7)\alpha_2^{-1}-(5/7)\alpha_1^{-1}=1/0.116$. 
On the other 
hand, since for $N_0=0$ the GUT scale $M_G$ is given approximately 
by $\ln{(M_G/M_Z)}\approx (5\pi /14)(\alpha_1^{-1}-\alpha_2^{-1})
\approx 33$ we have low scale unification if and only if
\begin{equation}
                      \hat{b}_2 < \hat{b}_1.
\end{equation}
Combining the last two equations we obtain 
\begin{equation}
                      \hat{b}_3< \frac{5}{8}\hat{b}_1+\frac{3}{8}\hat{b}_2,
\end{equation}
which, in view of the definition of $\kappa$, is nothing but the 
condition that $\kappa<1$. Further insight can be gained by using 
(\ref{eq:1}) to rewrite $\kappa$ as 
\begin{equation}
\kappa = \frac{8}{3}\frac{\alpha_3^{-1}(M_Z)+o_1}{\alpha^{-1}(M_Z)+o_2},
\end{equation}
where
\begin{equation}
                      o_1 \approx \frac{3}{2\pi}\ln\frac{M_G}{M_Z} -\Delta_3+\frac{\hat{b}_3}{4\pi}\ln{(M_G R)}
\end{equation}
\begin{equation}
                      o_2 \approx -\frac{6}{\pi}\ln\frac{M_G}{M_Z}-\frac{5}{3}\Delta_1-\Delta_2 + \frac{5\hat{b}_1+3\hat{b}_2}{12\pi}\ln{(M_G R)}
\end{equation}
are of order $O(1)$ in the case of low scale unification. While $o_2$ is 
much smaller than $\alpha^{-1}$, $o_1$ is still $\sim \log_{10}{(M_G/M_Z)}$ 
and therefore
\begin{equation}
\kappa \approx  \frac{8}{3}\frac{\alpha_3^{-1}(M_Z)}{\alpha^{-1}(M_Z)}
(1 \pm o)= 0.16 (1 \pm o),
\la{aprkapa}
\end{equation}
(for $\al_3(M_Z)=0.12$) where the model dependent quantity $o$ is
usually smaller than one 
as we will see in the next section.
{}From (\ref{aprkapa}) we already see that (\ref{kapconst}) can be
easily satisfied with $\kappa $ being near to the upper bound.
Therefore, a large number of KK states can suppress the 
relative variation of $\al_3$.

Finally, we want to emphasize that the introduction of more extra 
dimensions doesn't necessarily lead to a stronger suppression of the 
variation of $\alpha_3$. 
At least in the simple case, when the extra compact dimensions form a torus
of trivial shape and equal size radii,
the power law function $P_a/\hat{b}_a$ should be replaced by $(P_a/\hat{b}_a)^{\de
}$
($\de =$ number of extra dimensions). Limitations from scales require the 
$(P_a)^{\de }$ to have nearly the same value as $P_a$ in the case of $\de 
=1$. Therefore,
the relation between low scale unification and 
a suppressed $\delta\alpha_3$, as well as eq.\eqref{aprkapa}, remain 
unchanged. 
However, it is not excluded that with more elaborated compactifications 
a different behavior emerges.


\section{The Models}

In this section we present detailed models of low scale unification and
confirm the conclusions of the previous section.

\subsection{Low scale unified 5D SUSY $G_{321}$ models}

Consider the 5D SUSY $G_{321}$ model on an $S^{(1)}/Z_2$
orbifold. The
latter is crucial in order to have 4D $N=1$ SUSY at the orbifold fixed
point
(identified with our 4D world). As it will turn out, suppression of 
$\de \al_3/\al_3$ on the $M_Z$ scale occurs with a large number of KK
states
below the GUT scale $M_G$. Because of this, we consider specific
extensions \cite{kakushadze}, \cite{us} which naturally lead to low scale
unification.

a) In addition to gauge fields and $\eta $ matter families in the bulk, we
introduce the bulk states ${\bf E}_{N=2}^{(i)}= (E,~\ov{E})^{(i)}$ ($i=1, 2$),
 where the $E^{(i)}$, $\ov{E}^{(i)}$ are $SU(3)_c$,
$SU(2)_L$ singlets with $U(1)_Y$ hypercharges $6$, $-6$, resp., in 
$1/\sq{60}$ units. 
With orbifold parities for fragments 
$(E^{(1)},~\ov{E}^{(2)})\sim +$, $(E^{(2)},~\ov{E}^{(1)})\sim - $,
only $E^{(1)}$, $\ov{E}^{(2)}$ states will have zero modes with some
4D mass $M_E$.
With this setting, using (\ref{al3})-(\ref{alG}) we get
$$
\alpha_3^{-1}= \frac{12}{7}\alpha_2^{-1}-
\frac{5}{7}\alpha_1^{-1}+\frac{3}{7\pi}\ln{\frac{M_G}{M_E}},
$$
$$
\ln{\frac{M_G}{M_Z}}= \frac{5\pi}{14}(\alpha_1^{-1} - 
\alpha_2^{-1})-\frac{3}{14}\ln{\frac{M_G}{M_E}}-S,
$$
\beq
\alpha_G^{-1}=\alpha_2^{-1}-\frac{1}{2\pi}\ln{\frac{M_G}{M_Z}}+ 
\frac{1-2\eta}{\pi}S~, 
\la{ext1G321}
\eeq
where we have taken into account that 
$(b_1,~b_2,~b_3)^E=( \fr{6}{5}, ~0,~0)$, 
$(\hat{b}_1,~\hat{b}_2,~\hat{b}_3)=(\frac{18}{5},-2,-6)+4\eta\,(1,~1,~1)$.
Using (\ref{eq:kappa}), (\ref{ext1G321}), by straightforward
calculations one can verify that
with increase of $N_0$ the value of $\kappa $ decreases. However,
the increase
of $N_0$ is limited ($\stackrel{<}{_\sim }30$) from (\ref{ext1G321}) in
order to
have $M_G\stackrel{>}{_\sim }1$~TeV. From (\ref{ext1G321}) we see that we
can
have unification in the range $10$~TeV-$10^{16}$~GeV  varying $N_0$
between $0$ and $30$ and eqs. (\ref{ext1G321}) do not give a preferable
choice
for
$N_0$. However, if we want $\kappa $ suppressed, this
dictates large $N_0$($=30$) and therefore low scale
unification. More
precisely, for $N_0=30$, $M_E\simeq M_G$, 
$\mu_0=316$~GeV, $\eta =0$, we have
$M_G\simeq 10$~TeV
and $\kappa \simeq 0.2$. Since the contributions of 
$\ov{\rho }_c$, $\rho_a$ can provide some partial cancellations 
(which have not to occur with a high accuracy), this value of $\kappa $ can easily satisfy the bound \eqref{kapconst}.

b) A similar effect can be obtained by extending the 5D $G_{321}$ model
with
bulk states 
${\bf U^c}_{N=2}^{(i)}=(U^c, \ov{U}^c)^{(i)}$,
${\bf L}_{N=2}^{(i)}=(L, \ov{L})^{(i)}$ ($i=1, 2$), where 
$U^c$ and $L$ have $G_{321}$ quantum numbers
$(\ov{\bf 3}, {\bf 1}, 4)$ and $({\bf 1}, {\bf 2}, -3)$ resp. 
This extension also allows for low scale
unification. With orbifold parity prescriptions 
$({U^c}^{(1)}, {\ov{U}^c}^{(2)}, L^{(1)}, \ov{L}^{(2)})\sim +$,
$({U^c}^{(2)}, {\ov{U}^c}^{(1)}, L^{(2)}, \ov{L}^{(1)})\sim -$
we will have now 
$(\hat{b}_1,~\hat{b}_2,~\hat{b}_3)=(\frac{28}{5}, 0,-4)+4\eta\,(1,~1,~1)$.
Small values of $\kappa $ are still realized for large values of
$N_0$. Namely, 
for zero mode masses $\sim M_G$ of additional vector-like states and for 
$N_0=30$, $\eta=0$, $ \mu_0\simeq 316$~GeV unification holds at
$M_G\simeq 10$~TeV.
In this case we obtain $\kappa \simeq 0.21$. 

Let us note that within scenarios a) and b) it is possible to introduce
a pair of MSSM Higgs doublets on the brane (without KK excitations) and
make an
additional extension by introducing a vector like pair of $N=2$ SUSY
doublet
supermultiplets in the bulk with masses for zero modes near $M_G$. In
this case, the KK spectra will be precisely the same, RG analysis will not
be
altered and we still have $\kappa \simeq 0.2$. However, since in this case
all
matter and MSSM higgses have only brane couplings, their masses and
non gauge couplings will not depend on $R$ at tree level and therefore
are time independent at the leading
order. Because of this, the  assumptions made for our
estimates become selfconsistent.

\subsection{Low scale unified 5D SUSY 
$SU(4)_c\times SU(2)_L\times SU(2)_R$ model}

The other model we will present here has as its GUT symmetry the 
Pati-Salam gauge group $SU(4)_c\times SU(2)_L\times SU(2)_R$ \cite{pati}. 
By compactifying the 5th dimension on $S^1/Z_2\times Z_2'$ in a 
suitable way one obtains a 4D SUSY model (on a fixed point) with 
$SU(3)_c\times SU(2)_L\times SU(2)_R\times U(1)'$ gauge symmetry. 
Obviously, this is still not the symmetry of the standard model and we 
will therefore introduce a symmetry breaking 
$SU(2)_R\times U(1)'\to U(1)_Y$ by the Higgs mechanism at an intermediate 
scale. 
%
%
The field content consists of the 
minimal set of gauge and matter supermultiplets one needs to introduce 
to obtain the MSSM fields at low energies, a set of Higgs supermultiplets 
to break $SU(2)_R\times U(1)'$ at the intermediate scale, and a set of 
four 5D supermultiplets ${\bf R}^r$, doublets of $SU(2)_R$. This 
model (called III'-susy422 in ref.\cite{us} - see this ref. for
more details) allows low scale 
unification with $M_G \stackrel{>}{_\sim } 10^{5.8}$ GeV. 

Using the expressions given in \cite{us} for the gauge couplings 
$\alpha_a(M_G)$ (eqs. (7.64)-(7.66) of that reference) it is
straightforward 
to calculate $\kappa$ 
\begin{equation}
                 \kappa \simeq \frac{24-\frac{3}{2\pi}(N+1)}{24+\frac{9}{4\pi}(N+1)}.
\end{equation} 
Therefore, small $\kappa $ favors large $N$ and therefore low scale
unification. 
The phenomenological bounds \cite{bound} on the value of the intermediate
scale imply $N\leq 29$. For $N=29$ we obtain
$\kappa \simeq 0.2$ - a value which easily satisfies the bounds, as pointed out in the previous subsection.

\section{Conclusions}

We have shown that the extra-dimensional constructions provide not only
the
source for a variation of couplings but also offer possibilities of building
GUTs with phenomenological implications compatible with data. We have
demonstrated this in concrete 5D SUSY scenarios, which give
unification in multi TeV scales. A large number of KK states
below the GUT scale, plays an important role for the suppression of 
$\fr{\de \al_3(M_Z)}{\al_3(M_Z)}$ as well as for low scale
unification.

The change in the extra dimension's size is related to the radion dynamical
field, which should have a mass close to the Hubble parameter $\sim
10^{-33}$~eV (at $z=0.5-3.5$ redshift) in order to be relevant for a
variation
of gauge couplings \cite{dvali}.  In this way the radion can appear as a
quintessence field. All this crucially depends on the radion dynamics and
particularly on a mechanism through which the radion gets stabilized, but
still having small oscillations near its minimum.
Detailed studies of concrete examples are highly desirable as they allow to
discuss other possible cosmological and astrophysical implications
\cite{inprep}.

\vspace{1cm}
{\bf Acknowledgements}

We thank T. Dent and C. Grojean  for discussions 
and useful remarks. F.P.C. is supported 
by Funda\c c\~ ao para a Ci\^ encia e a Tecnologia (grant 
SFRH/BD/4973/2001).

\bibliographystyle{unsrt}

\end{document}